\documentclass[reprint,amsmath,amssymb,aps,prc,10 pt]{revtex4-2}
\usepackage{graphicx}
\usepackage{dcolumn}
\usepackage{bm}
\usepackage{amsmath}
\usepackage{multirow}
\usepackage{tabularx}
\usepackage[table]{xcolor}
\usepackage{lipsum}
\usepackage{blindtext}
\usepackage{natbib}

\begin{document}

\preprint{APS/701-Blocking}

\title{\textcolor{black}{On the timescale of quasi fission and Coulomb fission}}
\author{T. Nandi$^{1\dagger}$, H. C. Manjunatha$^{2\ddagger}$, P. S. Damodara Gupta$^2$, N. Sowmya$^2$,  N. Manjunatha$^2$, K. N. Sridhara$^2$, and L. Seenappa$^2$}
\affiliation{$^{1}$Inter University Accelerator Centre, Aruna Asaf Ali Marg, New Delhi 110067, Delhi, India} 
\affiliation{$^2$ Department of Physics, Government College for Women, Kolar, Karnataka, 563101 India}
\thanks{For correspondences:\\ $^\dagger$nanditapan@gmail.com \\ $^\ddagger$manjunathhc@rediffmail.com}

\date{\today}

\begin{abstract}

\textcolor{black}{Coulomb fission mechanism may take place if the maximum Coulomb-excitation energy transfer in a reaction exceeds the fission barrier of either the projectile or target. This condition is satisfied by all the reactions used for the earlier blocking measurements except one reaction $^{208}$Pb + Natural Ge crystal, where the measured timescale was below the measuring limit of the blocking measurements $<1~as$. Hence, inclusion of the Coulomb fission in the data analysis of the blocking experiments leads us to interpret that the measured time longer than a few attoseconds (about 2-2.5 $as$) is nothing but belonging to the Coulomb fission timescale and shorter than 1 $as$ are due to the quasifission. Consequently, this finding resolves the critical discrepancies between the fission timescale measurements using the  nuclear and blocking techniques. This, in turn, validates the fact that the quasifission timescale is indeed of the order of zeptoseconds in accordance with the nuclear experiments and theories. It thus provides a radical input in understanding the reaction mechanism for heavy element formation via fusion evaporation processes. } 
\end{abstract}

\maketitle
\begin{figure}
    \includegraphics[width=0.999\linewidth]{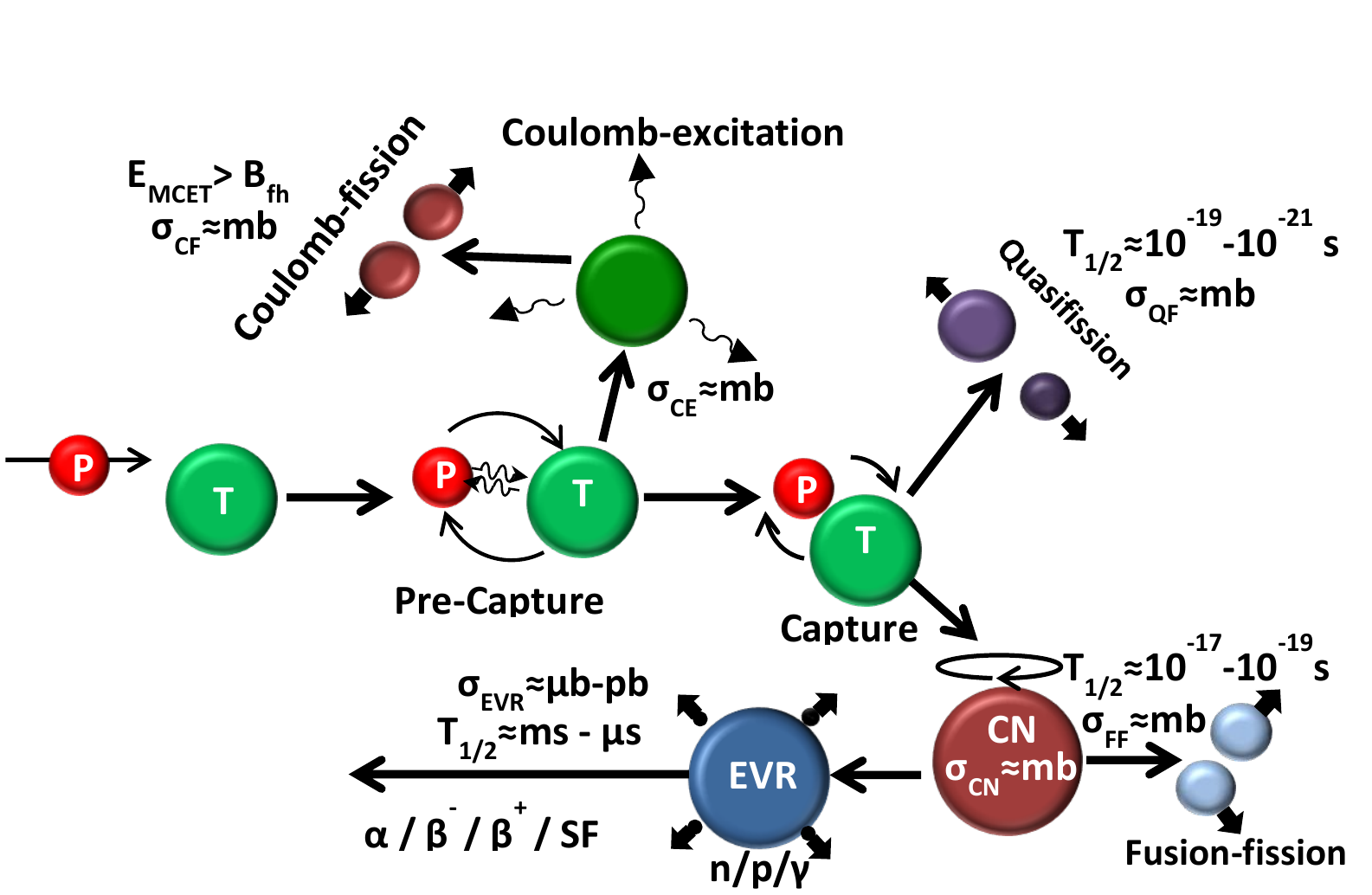}
    \caption{Synthesis of superheavy elements. P, projectile; T, target;  CN, compound nucleus; EVR, evaporation residue; SF, spontaneous fission. Possible processes such as Coulomb-fission followed by Coulomb-excitation in pre-capture stage, quasifission and formation of CN at post-capture stage, and fusion-fission and evaporation at post-CN stage. Finally,  $\beta^{-},  \beta^{+}, \alpha$  and SF decay at the post evaporation stage bring the reaction to a ground state. Thus, there are three potential outcomes from capture: Coulomb fission, quasifission and fusion. Various processes are characterized by corresponding cross-sections and half-lives. See further details in text.}
    \label{f1}
\end{figure}

The superheavy elements (SHEs) up to the element oganesson ($^{294}_{118}$Og) have been discovered that has completed up to the seventh row of the periodic table \cite{nazarewicz2018limits}. Currently, the synthesis of SHEs for Z $>$118 has been more significant and challenging field of nuclear research as initial experimental efforts to discover the elements $Z = 119$ and 120 of the eighth row have been failed at FLNR Dubna and GSI Darmstadt \cite{dullmann2016search,oganessian2009attempt,hofmann2015super}. The failure contradicts to the fact that the detailed properties of the nuclei synthesized at higher yields suggest increasing stability for superheavy elements Z$>$110 \cite{hamilton2013search,oganessian2015super}. It may be inferred that the evaporation cross-sections can be too low to measure. Now special effort is being invested to revamp the detection sensitivity close to fb. On the other side, we have taken up a project in pinning the possible reason of the debacle. One can think that the synthesis is strongly hindered by the competing faster processes such as quasifission (QF) because of fast splitting of the composite systems \cite{schroder1984damped}. To understand the process, a lot of effort has been put in the QF timescale measurements \cite{toke1985quasi,shen1987fission,hinde1992neutron,nestler1995time,velkovska1999quasifission,wilschut2004developing,goldenbaum1999fission,andersen2007crystal,andersen2008attosecond,morjean2008fission,ramachandran2006fission,du2011predominant,fregeau2012x} and theoretical studies \cite{siwek1995role,wilczynski1996determination,diaz2004modeling,zagrebaev2005unified,santhosh2017synthesis,manjunatha2017projectile,manjunatha2017survival,sridhar2018search,manjunatha2018investigations,sridhar2019studies,manjunatha2019investigations}. Though QF timescale measured by the nuclear techniques using the mass-angle distribution \cite{toke1985quasi,shen1987fission,velkovska1999quasifission,du2011predominant}, the neutron-clock method \cite{hinde1992neutron,ramachandran2006fission} and the giant dipole resonances \cite{nestler1995time} is in good agreement with the theoretical predictions \cite{siwek1995role,wilczynski1996determination,diaz2004modeling,zagrebaev2005unified,santhosh2017synthesis,manjunatha2017projectile,manjunatha2017survival,sridhar2018search,manjunatha2018investigations,sridhar2019studies,manjunatha2019investigations}, but it being at least two orders of magnitude lower than the values of the order of 10$^{-18}$s as measured by the atomic techniques using the crystal blocking \cite{goldenbaum1999fission,andersen2007crystal,andersen2008attosecond,morjean2008fission} and x-ray fluorescence \cite{wilschut2004developing,fregeau2012x}. The very long fission times can only be possible if the fission barriers of the isotopes involved in the decay chain are high, however theoretical estimates do not support it \cite{dobrowolski2009fission,manjunatha2018parameterization}. Further, recently a study \cite{sikdar2016examination} shows that short timescale $10^{-20}$ s from nuclear techniques and long  timescale from atomic techniques can only be reconciled by an extreme bimodal fission time distribution comprising a 53\% long-lived component (up to $10^{-16}$s) and a 47\% short-lived component (up to $10^{-21}$ s), provided the neutron emission from the accelerating fragments are  ignored. Hence, such an extreme bimodal distribution looks implausible. The authors thus conclude: "The incompatibility among the measured fission times by the nuclear and atomic techniques might indicate new physics \cite{ray2015quasifision} beyond the scope of fission physics."\\
\indent In this present work, we take an attempt to examine the methodologies of crystal blocking experiments \cite{goldenbaum1999fission,andersen2007crystal,andersen2008attosecond,morjean2008fission}. 
Besides the QF process, another belittling process against SHE synthesis may play decisive role, see Fig.\ref{f1}. It is known as Coulomb fission (CF) proposed in sixties \cite{wilets1967coulomb} and experimentally verified in late seventies \cite{backe1979direct}. The CF is a fission of a nucleus (see a review  \cite{oberacker1985coulomb}) induced by the time-varying Coulomb field of another nucleus passing by outside the range of the strong nuclear force, called Coulomb-excitation (see a classic review \cite{alder1956study}). Worth noting that Coulomb-excitation energy transfer ($\Delta E_{CET}$) can exceed the fission barrier of the actinide projectiles quite comfortably. For example,  $^{238}$U bombarded on $^{197}$Au at 24.3 MeV/u \cite{piasecki1996evidence,bonaccorso1997separation}. 
The CF is inevitable during the crystal blocking experiments in particular when the actinides are used in the reactions \cite{goldenbaum1999fission,morjean2008fission} because of low fission barrier of the actinides $~$6 MeV \cite{dobrowolski2009fission,manjunatha2018parameterization}. Even CF cannot be escaped for partners having high fission barrier used in the blocking experiment \cite{andersen2007crystal,andersen2008attosecond}, e.g., tungsten ($\approx$ 22 MeV).  Once we incorporate the CF along with QF and fusion-fission (FF) and analyse the experimental data thoroughly we find a discerning picture that the crystal blocking experiments do measure, in fact, the CF timescales. \\
\indent Prior to forming either the compound nucleus (CN) or dinuclear system (DNS), the projectile is slowed down once it enters within the interaction barrier zone and starts emitting electromagnetic radiation that can excite projectile or target nuclei, so called Coulomb excitation (CE). All these aspects (CN, DNS, CE) are governed by the nucleus-nucleus interaction potential, which includes the Coulomb, nuclear and rotational potentials. This potential along with the surface diffuseness parameter associated with the nuclear potential takes an important role in calculating various nuclear properties \\
\indent Normally, the compound nuclei are formed in the excited states, which undergo evaporation of neutrons (n), protons (P) and $\gamma$-photons. Evaporation residue cross-section ($\sigma_{EVR}$) for heavy element formation via the fusion evaporation can be evaluated by \cite{yanez2014measurement}:
\begin{align}
\sigma_{EVR} =\sum^{\infty}_{J=0} \sigma_J(E_{CM},J)P_{CN}(E^*,J)W_{sur}(E^*,J) \label{Eq9}
\end{align}
where  $\sigma_J(E_{CM},J)$ is the capture cross-section as a function of center-of-mass energy $E_{CM}$ and angular momentum J$\hbar$,  $P_{CN}(E^*,J)$ is the probability that the compound nucleus (CN) reaches the equilibrium configuration as a function of the excitation energy $E^*$ and J, and $ W_{sur}(E^*,J)$ is the probability that the completely fused system will deexcite by neutron emission rather than fission. Note that the complete fusion process is intensely hindered by quasifission (QF) and thus the probability P$_{CN}(E^*,J)$ \cite{hinde2002severe} is written as:
\begin{align}
P_{CN}(E^*,J)= 1 - P_{QF} \label{PCN1}
\end{align}
The above equation takes a central role in evaluating the fusion cross-section also. Detailed methodologies to obtaining the fusion and evaporation residue cross-section given in \cite{manjunatha2017survival,manjunatha2018investigations,sridhar2018search,manjunatha2019investigations,manjunatha2017survival} use \textcolor{black}{the theoretical mass excess tables} of Moller and Nix \cite{moller1995nuclear}.  Fusion-fission cross-sections are evaluated using a statistical model code HIVAP \cite{reisdorf1992well}, which uses the above said evaporation theory including the competitions between different decay modes of $\beta^-$, $\beta^+$, $\alpha$, and spontaneous fission (SF). Fission barriers have been obtained using the prescriptions given in \cite{cohen1963deformation,wong1973interaction,sierk1986macroscopic}. \\
\indent Coulomb-excitation processes occur in every nuclear reaction irrespective of the beam energy as portrayed in a schematic (Fig.\ref{f1}). We can calculate CET in terms of the energy of the incident projectile, charge, mass number, initial and final velocity, and excitation energy \cite{alder1956study, wollersheim2011relativistic}. To start with, the final projectile velocity is given by:
\begin{align}
     \frac{1}{2}m_1v_f^2=E_{lab}-\Delta E'\;\;\;{and}\; \Delta E'=\left(1+\frac{A_1}{A_2}\right)\Delta E. \label{ece3}
\end{align}
Where $A_1$ and A$_2$ are the nuclear mass number of projectile and target, $\Delta E$ is the excitation energy corresponding to a certain state of the rotational band of either projectile or target nuclei and  $\Delta E'$ is the excitation energy in lab frame. Introducing a dimensionless parameter
\begin{align}
    \chi=\frac{\Delta E}{E_{cm}}=\frac{\Delta E'}{E_{lab}}   \label{ece5}
\end{align}
and write the final velocity in terms of it as:
\textcolor{black}
{\begin{align}
    v_f=v_i(1-\chi)^{1/2} \;\;\;{where}\;\;\; v_i=\left(\frac{2E_{CM}}{\mu_0}\right)^{1/2}.   \label{ece6}
\end{align}
where $\mu_0$ is the reduced mass and $E_{CM}$ is the centre of mass energy of the projectile. Note that $\Delta E$ values are taken from \cite{nndc}. Using Eqns. 4 and 5, the classical symmetrized parameter \cite{alder1956study}, $ a = Z_1Z_2e^2/(\mu_0 v_i v_f)$, can be written as: 
\begin{align}
    a=\frac{Z_1Z_2e^2}{2 E_{cm}(1-\chi)^{1/2}}= \frac{0.7199 Z_1Z_2}{E_{cm} (MeV)(1-\chi)^{1/2}} fm \label{a}
\end{align}
where $Z_1$ and $Z_2$ are the charge of the projectile and target nuclei.} Let us define a basic parameter by the ratio of the collision time 
and the nuclear excitation time, 
the so-called dimensionless adiabaticity parameter \cite{alder1956study}, $\xi$, as:
\begin{align}
    \xi=\frac{0.07905 Z_1Z_2\mu_0^{1/2}\Delta E}{\left(E_{cm}-\frac{1}{2}\Delta E\right)^{3/2}} \left(1+\frac{5}{32}\left(\frac{\Delta E}{E_{cm}}\right)^2+...\right), \label{xi}
\end{align}
\noindent where both $\Delta E$  and $E_{cm}$ are in MeV. 
For low-energy collisions ($<$ 5 MeV/A), the $\Delta E_{CET}$ is expressed as:
\begin{align}
    \Delta E_{CET}= \hbar c \frac{ \beta \xi \gamma }{a-R} MeV;\beta = \frac{\sqrt{v_i v_f}}{c}, \gamma= \frac{1}{(1-\beta^2)^\frac{1}{2}}  
   \label{CET1}
\end{align}
 Where $\hbar$ is  Planck constant, c is speed of light and $R=\frac{1}{\sqrt{2}}\sqrt{R_1^2+R_2^2}$ and $R_i = 1.233 A_i^\frac{1}{3}-0.98 A_i^{-\frac{1}{3}} fm~(i=1,2)$ \cite{christensen1976evidence} correspond to radii of projectile and target nuclei. The $\Delta E_{CET}$ for  high energy collisions ($>$ 5 MeV/A)  is: 
\begin{align}
    \Delta E_{CET}= \hbar c \frac{ \beta \xi \gamma }{D-a} MeV; D=\frac{a}{\gamma}\times[1+arcsin(\theta^{gr}_{cm})] fm
    \label{CET2}
\end{align}
\noindent D is the distance of closest approach and $\theta^{gr}_{cm}$ is the grazing angle in radian as given by \cite{wilcke1980reaction}:
\begin{align}
\theta^{gr}_{cm}=2 arc sin\frac{\eta}{k_\infty \times R_{int}-\eta}; k_\infty=0.2187\mu_0\sqrt\frac{E_{lab}}{A_1}  \label{thgr}
\end{align}
Where the Sommerfeld parameter, $\eta=a k_\infty$ and $k_\infty$ is the reduced wave vector at infinite ion-target separation and the interaction radius, $R_{int}=A_1^{1/3}+A_2^{1/3}+4.59-\frac{A_1^{1/3}+A_2^{1/3}}{6.35}$ fm \cite{wilcke1980reaction}. 
Note that the maximum Coulomb-excitation energy transfer (MCET) is obtained by substituting $\xi=1$ in Eqn. \ref{CET1} and \ref{CET2}. While $\Delta$E is approximated from two neighboring $\Delta$Es around $\xi=1$ to estimate the $\chi$ as required for evaluating $a$ from Eqn.\ref{a}.\\
\begin{table*}
\caption{Examining the reactions used for blocking experiments \cite{andersen2007crystal,goldenbaum1999fission,morjean2008fission}. The reactions are specified by projectile (Proj), target (Targ) and projectile lab energy used (E$_{p}$), fusion barrier (V$_b$) \cite{manjunatha2018fusion}, evaporation channel (chnl) and fission-barrier of the compound nucleus (B$_{fc}$). Evaporation residue cross-section ($\sigma_{evr}$) \cite{reisdorf1992well}, fusion-fission lifetime ($\tau_{ff}$) \cite{nasirov2007angular}, quasifission barrier \cite{soheyli2016theoretical} and cross-section ($\sigma_{qf}$) \cite{tarasov2016lise++}, quasifission lifetime ($\tau_{qf}$) \cite{khanlari2017quasifission}, maximum Coulomb-excitation energy transfer (MCET), fission barrier of the heavier - either projectile or target nuclei (B$_{fh}$) \cite{sierk1986macroscopic}, Coulomb fission cross-section ($\sigma_{cf}$) \cite{tarasov2016lise++}, highest spin $I$ state that can be populated, and corresponding $\Delta E_{CET}$.}
\begin{tabular}{|c|c|c|c|c|c|c|c|c|c|c|c|c|c|c|c|c|}
\hline
Proj. &
  Targ. &
  \begin{tabular}[c]{@{}c@{}}E$_{p}$\\ (MeV)\end{tabular} &
  \begin{tabular}[c]{@{}c@{}}V$_b$\\ (MeV)\end{tabular} &
  Chnl &
  \begin{tabular}[c]{@{}c@{}}B$_{fC}$\\ (MeV)\end{tabular} &
  \begin{tabular}[c]{@{}c@{}}$\sigma_{EVR}$\\ (pb)\end{tabular} &
  \begin{tabular}[c]{@{}c@{}}$\tau_{ff}$\\ (as)\end{tabular} &
  \begin{tabular}[c]{@{}c@{}}B$_{qf}$\\ (MeV)\end{tabular} &
  \begin{tabular}[c]{@{}c@{}}$\sigma_{qf}$\\ (mb)\end{tabular} &
  \begin{tabular}[c]{@{}c@{}}$\tau_{qf}$\\ (zs)\end{tabular} &
  \begin{tabular}[c]{@{}c@{}}MCET\\ (MeV)\end{tabular} &
  \begin{tabular}[c]{@{}c@{}}B$_{fh}$\\ (MeV)\end{tabular} &
  \begin{tabular}[c]{@{}c@{}}$\sigma_{cf}$\\ (mb)\end{tabular} &
  I &
  \begin{tabular}[c]{@{}c@{}}$\Delta E_{CET}$\\ (MeV)\end{tabular} &
  Ref. \\ \hline
$^{32}S$   & $^{186}W$ & 170  & 136.8 & 2n & 8.2 & 7.3E5  & 14.3 & 10.24 & 4.5    & 50.5 & 46.1 & 21.4 & 0.1    & 8 & 19.4 &  \cite{andersen2007crystal} \\ \hline
$^{32}S$   & $^{184}W$ & 170  & 136.5 & 2n & 9.5 & 2.5E6 & 14.3 & 10.87 & 4.3    & 34.5 & 43.6  & 21.1 & 0.1    & 8 & 17.0 &  \cite{andersen2007crystal} \\ \hline
$^{48}Ti$  & $^{186}W$ & 250  & 183.9 & 3n & 2.2 & 4280 & 13.9 & 14.31 & 2.9    & 8.9  & 83.2  & 21.6 & 9E-2   & 4  & 11.5 &  \cite{andersen2007crystal} \\ \hline
$^{48}Ti$  & $^{184}W$ & 250  & 183.5 & 3n & 2.1 & 1350 & 17.4 & 14.2  & 3.1    & 8.5  & 74.6  & 21.2 & 9.2E-2 & 6  & 18.7 &  \cite{andersen2007crystal} \\ \hline
$^{58}Ni$  & $^{186}W$ & 315  & 233.1 & 3n & 0.6 & 2 & 13.8 & 14.62 & 5.3    & 5.3  & 48.7  & 21.4 & 6.9E-2 & 8  & 19.4 &  \cite{andersen2007crystal} \\ \hline
$^{58}Ni$  & $^{184}W$ & 315  & 232.5 & 3n & 0.5 & 1 & 40.9 & 14.8  & 4.9    & 5.2  & 45.6  & 21.1 & 6.8E-2 & 8 & 17.9 &  \cite{andersen2007crystal} \\ \hline
$^{238}U$  & $^{28}Si$ & 5712 & 144.2 & 6n & 1.1 & 1 & 32.3 & 2.33  & 1.1E-4 & 2.1  & 224.5   & 4.9  & 12.8 & 8   & 4.1  & \cite{goldenbaum1999fission} \\ \hline
$^{238}U$  & $^{29}Si$ & 5712 & 143.7 & 6n & 1.1 & 1& 32.3 & 1.59  & 1.1E-4 & 1.9  & 242.2   & 4.9  & 12.5   & 8  & 4.2  & \cite{goldenbaum1999fission} \\ \hline
$^{238}U$  & $^{30}Si$ & 5712 & 143.2 & 6n & 0.7 & 1& 31.8 & 1.59  & 1.1E-4 & 1.2  & 257.8   & 4.9  & 12.2   & 8  & 4.4  & \cite{goldenbaum1999fission} \\ \hline
$^{238}U$  & $^{58}Ni$ & 1575 & 280.3 & 4n & 0.3 & 17& 43.7 & 1.7   & 4.5    & 2.9  & 4.2   & 4.0  & 1.6    & 24 & 3.8  & \cite{morjean2008fission} \\ \hline
$^{238}U$  & $^{60}Ni$ & 1575 & 279.1 & 4n & 0.3 & 9 & 43.2 & 1.59  & 4.3    & 3.4  & 4.7   & 4.1  & 1.5    & 22 & 3.6  & \cite{morjean2008fission} \\ \hline
$^{238}U$  & $^{61}Ni$ & 1575 & 278.5 & 4n & 0.5 & 2& 42.8 & 1.7   & 4.2    & 3.9  & 5.0   & 4.1  & 1.5    & 22 & 3.8  & \cite{morjean2008fission} \\ \hline
$^{238}U$  & $^{62}Ni$ & 1575 & 277.9 & 4n & 0.3 & 1& 42.5 & 1.7   & 4.1    & 4.6  & 5.3   & 4.1  & 1.5    & 22 & 3.9  & \cite{morjean2008fission} \\ \hline
$^{238}U$  & $^{64}Ni$ & 1575 & 276.8 & 5n & 0.4 & 20& 42.1 & 1.59  & 3.9    & 4.1  & 5.8   & 4.2  & 1.4    & 20 & 3.6  & \cite{morjean2008fission} \\ \hline
$^{238}U$  & $^{70}Ge$ & 1450 & 316.2 & 4n & 0.5 & 1& 41.0   & 2.23  & 3.2    & 4    & 3.3   & 3.2  & 1.2    & 26 & 3.1  & \cite{morjean2008fission} \\ \hline
$^{238}U$  & $^{72}Ge$ & 1450 & 314.9 & 4n & 0.1 & 1& 41.3 & 1.38  & 3.9    & 4.2  & 3.7   & 3.6  & 1.0      & 24 & 3.5  & \cite{morjean2008fission} \\ \hline
$^{238}U$  & $^{73}Ge$ & 1450 & 314.4 & 5n & 0.2 & 1& 41.2 & 1.27  & 3.9    & 4.4  & 3.8   & 3.7  & 0.9    & 24 & 3.6  & \cite{morjean2008fission} \\ \hline
$^{238}U$  & $^{75}Ge$ & 1450 & 313.2 & 2n & 0.2 & 1& 40.9 & 1.27  & 3.8    & 4.8  & 4.1   & 4.0  & 0.9    & 24 & 4.0  & \cite{morjean2008fission} \\ \hline
$^{238}U$  & $^{76}Ge$ & 1450 & 312.7 & 6n & 0.1 & 1& 40.7 & 1.38  & 3.6    & 4.8  & 4.3   & 4.1  & 0.8    & 22 & 3.6  & \cite{morjean2008fission} \\ \hline
$^{208}Pb$ & $^{70}Ge$ & 1281 & 286.3 & 3n & 0.3 & 3 & 2505 & 19.05 & 3.3    & 10   & 4.5  & 24.9 & 2.8E-8 & 16 & 20.0 & \cite{morjean2008fission} \\ \hline
$^{208}Pb$ & $^{72}Ge$ & 1281 & 285.2 & 3n & 0.3 & 1& 2430 & 18.17 & 3.5    & 7.8  & 4.9  & 24.9 & 1.6E-8 & 16 & 21.3 & \cite{morjean2008fission} \\ \hline
$^{208}Pb$ & $^{73}Ge$ & 1281 & 284.6 & 4n & 0.5 & 39& 2408 & 18.17 & 3.4    & 5.1  & 5.1  & 24.9 & 1.3E-8 & 16 & 21.9 & \cite{morjean2008fission} \\ \hline
$^{208}Pb$ & $^{75}Ge$ & 1281 & 283.6 & 4n & 0.5 & 49 & 2344 & 18.39 & 3.3    & 3.7  & 5.5  & 24.9 & 1.9E-8 & 16 & 23.3 & \cite{morjean2008fission} \\ \hline
$^{208}Pb$ & $^{76}Ge$ & 1281 & 283.1 & 4n & 0.4 & 58& 2308 & 18.39 & 3.2    & 7.9  & 5.8  & 24.9 & 1.8E-8 & 16 & 23.8 & \cite{morjean2008fission} \\ \hline
\end{tabular}
\label{data}
\end{table*}
\indent We have examined the reactions used for blocking experiments \cite{goldenbaum1999fission,andersen2007crystal,andersen2008attosecond,morjean2008fission} in Table I. Notice here that fusion-fission lifetimes ($\tau_{ff}$) \cite{nasirov2007angular} are $>$ 13 $as$, whereas the quasifission lifetime ($\tau_{qf}$) \cite{khanlari2017quasifission}  are estimated to be $<$ 50 zs. Hence, both the lifetimes are far from the precisely measured values 2.0-2.5 $as$. Obvious question arises: what is then measured in the blocking experiments? Comparison of maximum Coulomb-excitation energy transfer (MCET) with fission barrier ($B_f$) of  the heavier partners shows that the former is higher than the latter at the beam energy used except the reactions $^{208}$Pb + Ge single crystal. Having this condition satisfied, the reaction may exhibit the Coulomb fission (CF) and the fragments being detected in the blocking measurements having 2-2.5 $as$ lifetime. Whereas no CF fragments available in the reactions $^{208}$Pb + Ge single crystal and thus the measured fragments with timescale $<1 as$ must be ascribed undoubtedly to the QF process.\\
\begin{table}
\caption{Testing correctness of MCET calculation. 
Reactions are specified by projectile (Proj), Target (Targ), and projectile lab energy (E$_p$). MCET, fission barrier (B$_f$) of either projectile or target nucleus, estimated highest spin state $I$ and experimentally observed state $I^\prime$. The excited nucleus is denoted by *. }
\begin{tabular}{|c|c|c|c|c|c|c|c|c|}
\hline
Proj. &
  Targ. &
  \begin{tabular}[c]{@{}c@{}}E$_p$\\ (MeV)\end{tabular} &
  \begin{tabular}[c]{@{}c@{}}MCET\\ (MeV)\end{tabular} &
  \begin{tabular}[c]{@{}c@{}}B$_f$\\ (MeV)\end{tabular} &
  I &
  \begin{tabular}[c]{@{}c@{}}$\Delta E{CET}$\\ (MeV)\end{tabular} &
  I' &
  Ref. \\ \hline
$^{40}Ar$    & $^{232}Th^*$ & 158 & 6.7  & 6.6  & 16 & 5.9  & 10 & \cite{stephens1959multiple} \\ \hline
$^{40}Ar$    & $^{232}Th^*$ & 190 & 15.3  & 6.8  & 12 & 6.5  & 10 & \cite{stephens1959multiple} \\ \hline
$^{40}Ar$    & $^{238}U^*$  & 190 & 14.3  & 4.7  & 10 & 4.2  & 12 & \cite{stephens1959multiple} \\ \hline
$^{58}Ni$    & $^{248}Cm^*$ & 260 & 6.6  & 2.8  & 10 &  2.4 & 16 & \cite{czosnyka1986e2} \\ \hline
$^{136}Xe$   & $^{248}Cm^*$ & 641 & 9.5  & 2.8  & 8  & 2.2  & 22 & \cite{czosnyka1986e2} \\ \hline
$^{116}Sn$   & $^{162}Dy^*$ & 637 & 90.7   & 29.2 & 8  & 27.5 & 10 & \cite{wu1990pairing} \\ \hline
\end{tabular}
\end{table}
\begin{figure}
        \includegraphics[width=0.99\linewidth]{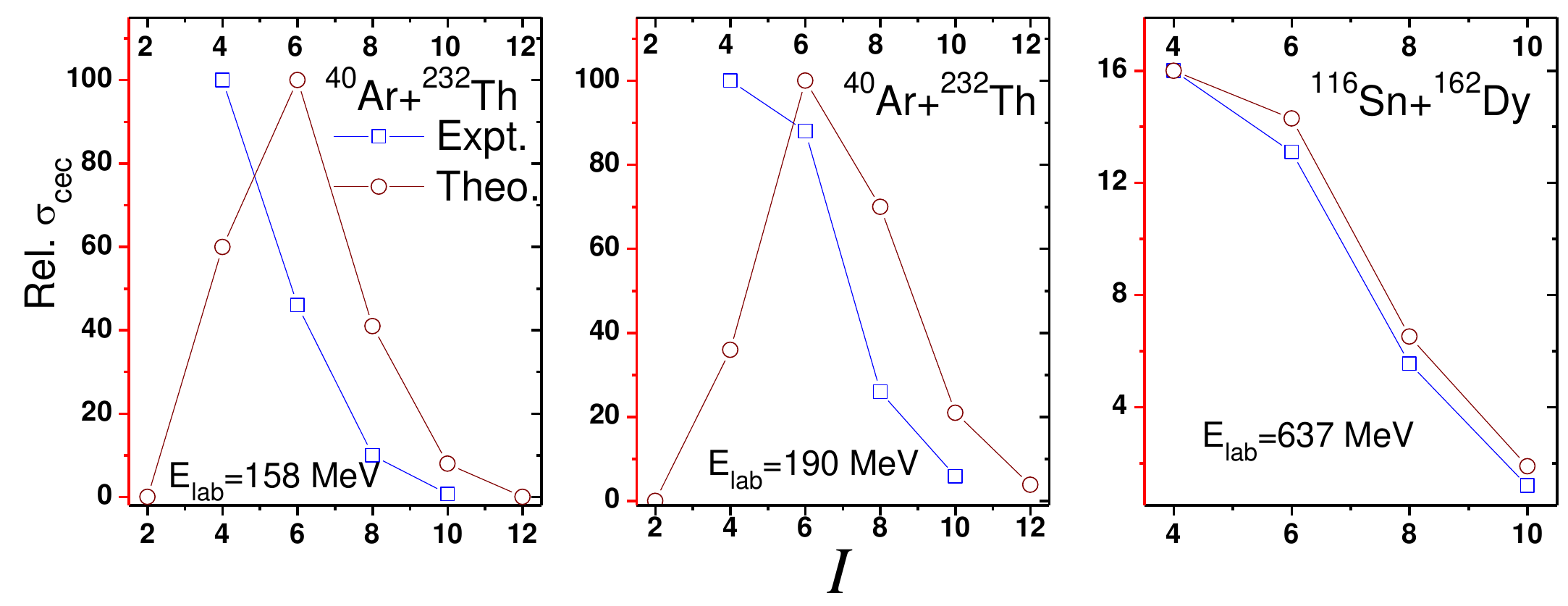}
        \caption{Relative Coulomb-excitation cross-section ($\sigma_{cec}$) vs spin $I$. Theoretical calculations are done using COULEX code \cite{winther1965computer}. Fig. (a) shows the data for $^{40}Ar + ^{232}Th$ reaction at 158 MeV: the maximum experimental $\sigma_{cec}$ is found at $I=4$, but the  theoretical counterpart is at $I=6$ and major trend is similar to each other. However, theory predicts up to $I=16$ ought to be observed, but not observed so because of too low cross-section to be measured for $I>10$, Fig. (b) shows the data for $^{40}Ar + ^{232}Th$ reaction at 190 MeV: the maximum experimental $\sigma_{cec}$ is found at $I=4$, but the  theoretical counterpart is at $I=6$, but major trend is similar here also. Fig. (c) shows the data for $^{116}Sn + ^{162}Dy$: the experiment and theory agree very well till $I=10$, but transitions for $I>10$ are also observed due to the large cross-section and large mean lifetime for Coulomb fission. See further details in text.}
    \label{f2}
\end{figure}
\indent To justify truth of the above mentioned CF mechanism, we have compared the highest spin state $I$,  populated before exceeding the fission barrier, estimated using Eqn. \ref{CET1} or \ref{CET2},  with experimentally observed values and shown in Table II. The reactions  $^{40}Ar + ^{238}U/^{232}Th$ at 190 MeV \cite{stephens1959multiple} and  $^{116}Sn + ^{162}Dy$ show good agreement. While the reaction $^{40}$Ar + $^{232}$Th at lower beam energy 158 MeV, experimental spin $I$ is 10$^+$, but theoretical value is 16$^+$. This difference can be attributed to too low Coulomb-excitation cross-section (CEC) for $I > 10$ to be measured in the experiment as shown in Fig.\ref{f2}.  The CEC is calculated using the COULEX code \cite{winther1965computer} and B(E2$\uparrow$) value of every participating nucleus is taken from NNDC site \cite{nndc}. In contrast, the experimental $I$ is higher than the theoretical limit in some cases; viz., $^{58}Ni/^{136}Xe$ + $^{248}Cm$ reactions. This can happen because of high CEC beyond fission barrier and long timescale of the CF process ~2.5 $as$. Note that the experimental CEC values peak at I=$4^+$, but the Coulex code predictions occur at $6^+$ for the reactions  $^{40}Ar + ^{238}U/^{232}Th$. Whereas relative experimental CEC from 4$^+$ to 10$^+$ is almost a replica of the theoretical curve from 6$^+$ to 12$^+$. \\
\indent \textcolor{black}{The QF lifetimes are evaluated as follows
\begin{align}
    \tau_{qf}=\frac{1}{\lambda_{qf}}
    \end{align}
    where $\lambda_{qf}$ is the QF decay constant and is expressed as
    \begin{align}
        \lambda_{qf}=&\frac{\omega_m}{2\pi\omega_{qf}}\left(\sqrt{\left(\frac{\Gamma}{2\hbar}\right)^2+\omega_{qf}^2}-\frac{\Gamma}{2\hbar}\right)\\\nonumber&\exp\left(-\frac{B_{qf}(Z,A,\ell)}{\Theta(Z,A)}\right).
    \end{align}
Here, the quantity ${\Gamma}$ denotes an average width of the contributing single-particle states near the Fermi surface and normally, its value is taken as 2 MeV. The quantities $\omega_m$ and $\omega_{qf}$ are the frequency of the harmonic oscillator and inverted harmonic oscillator, respectively and these have been evaluated using a set of equations given in \citet{manjunatha2021quasifission}. 
The nuclear temperature  $\Theta(Z,A)$ depending on the level density parameter $a$ is given by
\begin{align}
    \Theta(Z,A)=\sqrt{\frac{E^*(l)}{a}} \label{eq34}
\end{align}
where $E^*$ is the excitation energy imparted to the compound nucleus. The QF barrier $B_{qf}(Z,A,\ell)$ in the dinuclear system model is expressed as
\begin{align}
    B_{qf}(Z,A,l)=\nonumber&V(R_b,Z,A,\beta_{21},\beta_{22},l)-\\&V(R_m,Z,A,\beta_{21},\beta_{22},l). 
\end{align}    
Where $\ell$ and $R=R_m$ \cite{manjunatha2021quasifission} are the angular momentum and distance at which the nucleus-nucleus potential is minimum, respectively. $\beta_{21}$ and $\beta_{22}$ are the quadruple deformation parameters. The nucleus-nucleus interaction potential is given by
\begin{align}
V(R,Z_1,Z_2,\beta_{2i},l)=\nonumber& V_c(R,Z_1,Z_2,\beta_{2i})+V_{rot}(l,\beta_{2i}) \\&V_N(R,Z_1,Z_2,\beta_{2i}). 
\end{align}
Where $V_C$, $V_N$ and $V_{rot}$ are the Coulomb, nuclear and rotational potentials, respectively and are evaluated using a set of equations given by  Soheyli and  Khanlari \cite{soheyli2016theoretical}.\\
\indent The fusion-fission (FF) timescales are evaluated as follows
\begin{align}
    \tau_{ff}=\frac{1}{\lambda_{ff}}
\end{align}
where $\lambda_{ff}$ is the decay constant of the fusion-fission process and it is expressed as;
\begin{align}
    \lambda_{ff}=&\nonumber\frac{1}{2\pi}\left(\sqrt{\left(\frac{\Gamma_0}{2\hbar}\right)^2+\omega_f^2}-\frac{\Gamma_0}{2\hbar}\right)\times\\&\exp\left(-\frac{B_{f}(Z,A,\ell)}{\Theta(Z,A)}\right).
\end{align}
Here, $\Gamma_0=2$ MeV. The decay constant is taken from \citet{khanlari2017quasifission}. \\
\indent The following equation is used to  evaluate the $QF$ cross sections \cite{nasirov2013main} using the LISE$^{++}$ code \cite{tarasov2016lise++}.
\begin{align}
    \sigma_{qf}(E_{cm},\beta_{p},\alpha_{2})=&\nonumber\sum_{\ell=\ell_{f}}^{\ell_{d}}(2\ell+1)\sigma_{cap}(E_{cm},\ell,\beta_{p},\alpha_{2})\\&(1-P_{CN}(E_{cm},\ell,\beta_{p},\alpha_{2}))
\end{align}
where $\ell$ is the angular momentum, $p_{CN}$ is the compound nucleus formation probability and $\sigma_{cap}$  is the capture cross section which is the sum of quasifission, fusion fission and fast fission cross sections. \\
\indent Coulomb fission cross section  \cite{bonaccorso1997separation} is expressed below and evaluated using the LISE$^{++}$ code \cite{tarasov2016lise++}.
\begin{align}
    \sigma_{cf}=\int [n_{E_{1}}(\omega)\sigma^{E_{1}}(\omega)+n_{E_{2}}(\omega)\sigma^{E_{2}}(\omega)]P_{f}(\omega)\frac{d\omega}{\omega}
\end{align}
here $n_{El}(\omega)$ is the virtual photon spectra for the electric ($El$) multi polarities. Only excitation multipolarities necessary to include are $E_1$ and $E_2$. $\sigma^{E_{1}}$ and $\sigma^{E_{2}}$ are the photo absorption cross section for dipole and quadrupole, respectively. $P_{f}(\omega)$ is the fission probability \cite{grange1980fission}. \\
\indent The fusion barrier is evaluated using the empirical relation based on the experiments and it is given as \cite{nandi2020search}
\begin{align}
    B_{fu}=&\nonumber-34488.7618+1100.6666z-14.4066z^{2}+9.9275\\&\nonumber\times10^{-2}z^{3}-3.7959\times10^{-4}z^{4}
    +7.6357\times10^{-7}z^{5}\\&-6.3136\times10^{-10}z^{6}\;\mbox{for}\;128 \leq z  \leq 286.
\end{align}
Here, $z=\frac{Z_{1}Z_{2}}{A_{1}^{1/3}+A_{2}^{1/3}}$ is called Coulomb interaction parameter, where $Z_{1}$ and $Z_{2}$ are the atomic number and  $A_{1}$ and $A_{2}$ are the mass number of the projectile and target, respectively.}\\
\indent Both QF and CF cross-sections are found to be of the order of mb. At low excitation energies the QF cross-sections are larger than the CF cross-sections. The scenario takes a U-turn at high excitation energies \cite{bonaccorso1997separation}. Furthermore, Table I displays that the CF cross-section is small and the QF cross-section is large if the fission barrier of the heavier partner (B$_{fh}$) \cite{sierk1986macroscopic} is large irrespective of the excitation energy, e.g.,$^{32}S/^{48}Ti/^{58}Ni$ + $^{184,186}W$. If both the CF and QF cross-sections are large,  viz., $^{238}U + ^{58,64}Ni/^{70-73,75-76}Ge$, it will be difficult to distinguish between them in a measurement. But large differences in timescale may come into rescue as discussed later. In contrast, the experiment with the reactions $^{238}U + ^{28-30}Si$ performed at very high excitation energy shows the CF cross-section is four orders of magnitude higher than that of the QF cross-section. Hence, the experiment mostly measures the CF fragments and thus the measured timescales around 2.5 $as$ can only be attributed to the CF process.\\
\indent The MCET for the reactions $^{208}Pb + ^{70-73,75-76}Ge$ (see Table I) are far below the fission barrier of $^{208}$Pb and consequently the CF cross-sections are found to be eight orders of magnitude lower than that of QF. Hence, with this reaction, the blocking experiment can only detect the QF fragments and thus the timescale measurement can only be attributed to the QF timescales. This fact is further corroborated by the experimental outcome \cite{morjean2008fission} that these reactions exhibit the shorter lifetimes, below the sensitivity limit of the blocking experiment ($1 as$) \cite{morjean2008fission}.\\
\indent The QF and CF cross-sections have relevance with the evaporation and fusion-fission cross-sections. If former cross-sections are large, the latter cross-sections are small, which is undesirable for the synthesis of the SHEs. The challenge is to minimise the QF and CF cross-sections and maximise the evaporation and fusion-fission cross-sections. Nevertheless, the evaporation and fusion-fission move hand-in-hand. This is also unwanted, as we desire evaporation being higher, the fusion-fission cross-section can be minimised by a larger fission barrier of the compound nucleus that depends on the choice of the target-projectile combination for any reactions \cite{sridhar2018search,manjunatha2018investigation}. \\ 
\indent If the timescale measured in the blocking experiments signifies the CF characteristics, time measurements using the x-ray fluorescence \cite{wilschut2004developing,fregeau2012x} may also be representing the same CF process or something else. This is a crucial point to be divulged out to troubleshoot the fission timescale fully. This study is now under progress.\\
\indent We will discuss now a radical consequence of the present work. Fig.\ref{f1} shows three potential outcomes from pre-capture and capture: Coulomb fission, quasifission, and fusion. Furthermore, we have revealed that the CF plays  also a hindering role as the QF in the heavy ion reactions meant for forming the SHEs. Thus, the probability P$_{CN}(E^*,J)$ given in Eqn.\ref{PCN1} can be rewritten as:
\begin{align}
P_{CN}(E^*,J)= (1 - P_{QF})(1 - P_{CF})
\label{PCN2}
\end{align}
\noindent Inclusion of Eqn.\ref{PCN2} in the dynamical models may drastically reduce the $P_{CN}$. Consequently, this equation will assist in planning the synthesis of a wider range of superheavy isotopes, in particular, to choose the beam energy such that the CF is avoided.\\
\indent In summary, the Coulomb fission mechanism of the projectile/target nucleus at the pre-capture stage is very significant for any nuclear reactions. Its cross-section is comparable to that of quasifission occurring at the post capture stage. The Coulomb fission has led us to explain well the measured timescale by the blocking technique. It thus reveals that the attosecond timescale so measured belongs to the timing property of the Coulomb fission. On the other hand, it validates the zeptosecond timescale of the quasifission. All the blocking experiments have led to about the same Coulomb fission timescale $\approx$2.5 as. Now it is a challenge for the experiments as well as theories whether it is so or it may vary with varieties of reactions and experimental conditions such as entrance channel and beam energy. Moreover, we notice that the CF has a significant role in the CN formation probability, hence another challenge of the experiments and theories is to consider the CF and QF process on equal footing for the production of the superheavy elements. \\
\indent We have explored another dimension of the blocking method to investigate the Coulomb fission timescales. Though it indicates the presence of the QF by measuring the corresponding timescale shorter than an $as$ (the measurement limit of the technique), it cannot measure the exact figure as showcased by the reactions of $^{208}$Pb + Ge single crystal. In contrast, the nuclear techniques have the potential to assess perhaps both the Coulomb fission and quasifission in a single experiment, which is yet to be explored in an experiment. Last but not the least, we suggest to detect the Coulomb excitation spectra for every experiment involving the superheavy nucleus formation.\\
\indent We are extremely grateful to Professor H.J. Wollersheim for stimulating discussions on the physics of the Coulomb excitation and calculation of the Coulomb excitation cross-sections  using the Coulex code.\\
\bibliographystyle{apsrev4-2}
\bibliography{apssamp.bib}
\end{document}